\documentclass[rnote]{aa} 
\usepackage{graphicx}
\usepackage{txfonts}
\begin{document}
   \title{The HI properties of galaxies in the Coma I cloud revisited}


   \author{A. Boselli\inst{1}
           \and
 	   G. Gavazzi\inst{2}
          }

   \offprints{A. Boselli}

   \institute{Laboratoire d'Astrophysique de Marseille, UMR 6110 CNRS, 38 rue F. Joliot-Curie, F-13388 Marseille France\\
              \email{Alessandro.Boselli@oamp.fr}
	 \and
	     Universita degli Studi di Milano-Bicocca, Piazza delle Scienze 3,
	     20126 Milano, Italy
	     \email{Giuseppe.Gavazzi@mib.infn.it}
             }

   \date{}

 
  \abstract
    {Pre-processing within small groups has been proposed to explain 
    several of the properties of galaxies inhabiting rich clusters.}  
    {The aim of the present work is to see whether pre-processing is acting in the nearby universe, where the structures that are merging to form 
    rich clusters are rather large and massive.}
    {We study the HI gas properties of a large sample of late-type galaxies belonging to the Coma I cloud, an association of 
    objects close to the Virgo cluster.}
    {Contrary to what previously claimed, late-type galaxies in the Coma I cloud are not deficient in HI gas ($HI-def$=0.06$\pm$0.44).}
    {If the Coma I cloud is representative of infalling groups in nearby clusters, 
    this result suggests that, in the local universe, the evolution of late-type galaxies belonging to loose structures with 
    high velocity dispersions ($\geq$ 300 km s$^{-1}$)
    associated to rich clusters such as Virgo is not significantly perturbed by pre-processing. }


   \authorrunning{Boselli \& Gavazzi}
   \titlerunning{The Coma I cloud revisited}
   \maketitle

\section{Introduction}

The morphology segregation effect (Dressler 1980; Whitemore \& Gilmore. 1991)
is the strongest evidence that the environment played a major role in shaping galaxy evolution.
Recent surveys such as SDSS (Gomez et al. 2003) and 2dF (Lewis et al. 2002), which allowed to 
continuously trace galaxy properties from the highest 
density regions in the core of rich clusters down to the field, have shown that 
the star formation activity already decreases at the periphery of clusters, probably because 
the interactions responsible for removal of gas, the principal feeder of star formation (e.g. Boselli et al. 2001), 
were already in place in the infalling groups 
prior to the formation of rich clusters. These results are consistent with our own studies of the gas and star formation properties 
of galaxies in nearby clusters (Gavazzi et al. 2002; 2005; 2006a; 2006b; 2008).
Although in some cases the presence of hot gas might trigger galaxy interactions with
the intergalactic medium, the low velocity dispersion of small groups ($\leq$ 200 km s$^{-1}$) suggests that
gravitational interactions are probably at the origin of the pre-processing of galaxies even before they enter rich clusters
(Dressler 2004).\\
Probably efficient at high redshift, when clusters were under formation (Gnedin 2003),
pre-processing is less evident in the nearby universe (Boselli \& Gavazzi 2006), where clusters 
are rather accreting large structures characterized by high velocity dispersions 
(Donnelly et al. 2001; Ferrari et al. 2003; Cortese et al. 2004) or single galaxies, making 
gravitational interactions rather rare. For instance the velocity dispersion of the M and W clouds in the Virgo cluster
are relatively high, of the order of 450-650 km s$^{-1}$ (Gavazzi et al. 1999), thus almost comparable to that of an already formed cluster. 
The only exception found in the local universe is
the blue infalling group in A1367 (Sakai et al. 2002; Gavazzi et al. 2003a; Cortese et al. 2006), a compact group of galaxies
with a velocity dispersion of only 150 km s$^{-1}$ infalling into the cluster A1367. Here pre-processing is efficiently perturbing 
the galaxies morphology and star formation activity, creating long tails of ionized gas.\\
The study of the Virgo cluster, the nearest rich cluster of galaxies, and its surroundings, however, revealed the presence of 
satellite clouds with HI-deficient objects witnessing an ongoing interaction, thus making these clouds of particular interest 
for studying pre-processing in the nearby universe. Being loosely anchored to the galaxy potential, the HI component can be
easily removed during any kind of interaction, and is thus an ideal tracer of ongoing perturbations (Boselli \& Gavazzi 2006).
Among these, the Coma I cloud, a loose aggregation of galaxies
in the projected direction of the Coma/A1367 supercluster located at $\sim$ 5 Mpc from M87 (see sect. 4 and 5), 
is the most promising since
previous studies have shown that this loose cloud is composed of HI-deficient galaxies (Garcia-Barreto et al. 1994).
The availability of new HI data more than doubled the sample of Garcia-Barreto et al. (1994), suggesting us 
to reanalyze the HI gas properties of the Coma I cloud galaxies in the framework of pre-processing
in the nearby universe.

\section{The sample}

The Coma I cloud has been defined by Gregory \& Thompson (1977) as the cloud of nearby galaxies ($\le$ 20 Mpc) in the foreground
of the Coma/A1367 supercluster. 
The analysis presented in this work is thus based on a sample composed of all galaxies extracted from NED in the
sky region 11$^h$30$^m$ $<$ R.A.(2000) $<$13$^h$30$^m$; 20$^o$ $<$ dec $<$ 34$^o$ with a recessional velocity $\leq$ 2000 km s$^{-1}$. 
Excluding misclassified HII regions associated to bright galaxies, the resulting sample is composed of 
161 galaxies. Since no limits on the magnitude or diameter of the selected galaxies are applied, the selected 
sample is not complete in any sense. \\

\section{The data}

The set of data necessary for the following analysis, restricted to those galaxies with available HI data (72 objects), are listed in Table 1. 
This includes morphological type, optical and near IR magnitudes,
optical diameters and HI flux and line width measurements.
Coordinates and morphological type have been taken from NED, in its updated version including the SDSS data release 6.
For galaxies without a morphological classification, we assigned a morphology type according to, in order of preference, the presence of
emission lines in the SDSS spectra, their optical color on the SDSS composed image or their optical appearance on the POSS plates.
Thanks to their proximity, the morphological classification of the brightest galaxies, those with available HI data, thus the most
concerned by the present analysis, 
is very accurate, of less than one bin in the Hubble sequence. It is poor for the very compact sources that dominate
at low luminosity. The discrimination between early and late-type, based on spectroscopic measurements
or optical colors, however, should be credible.\\
Optical diameters have been taken form GOLDMine (Gavazzi et al. 2003b) whenever available, or from NED otherwise. We preferred 
to use the GOLDMine values whenever possible to be as consistent as possible in the definition and in the determination of the
HI-deficiency parameter, which is here based on the calibrations of Solanes et al. (1996). Near infrared JHK total magnitudes have been
taken from 2MASS (Jarrett et al. 2003) whenever available, or from Gavazzi \& Boselli (1996). The comparison of
2MASS and GOLDMine total magnitudes that we made using an extended sample indicates that they differ by less than 0.1 mag.\\
HI data have been taken from several sources in the literature: to be consistent with the distance determination using the 
Tully-Fisher calibration given by Masters et al. (2008), HI fluxes and velocity widths have been taken 
whenever available from Springob et al. (2005). These data have also the advantage of being accurately homogenized.
HI line widths from other sources have been corrected for smoothing, redshift stretch and turbulent motion (6.5 km s$^{-1}$) 
as prescribed in Springob et al. (2005). Consistently HI fluxes of sources others than Springob et al. (2005)
have been corrected for pointing offsets and beam attenuation. 
The accuracy in the fluxes should be of the order of 10-15\% (Springob et al. 2005),
and of the order of 10 km s$^{-1}$ in the line width measurements.\\
Galaxies in Table \ref{Tabdata} are arranged as follow:\\
Column 1: galaxy name.\\
Column 2: morphological type, from NED whenever available, or from our own classification.\\
Column 3 and 4: major and minor optical diameters, from GOLDMine whenever available, from NED elsewhere. These are
B band isophotal diameters at the 25 mag arcsec$^{-2}$.\\
Column 5: optical (generally B band ($m_B$)) magnitude, from NED.\\
Column 6: heliocentric velocity, in km s$^{-1}$, from NED.\\
Column 7: Tully-Fisher distance, in Mpc, determined as explained in sect. 4.2. Distances for those galaxies with available
primary indicators are taken from Ferrarese et al. (2000). For galaxies without a direct distance estimate, we assume 14.52 Mpc.\\ 
Column 8: Coma I cloud (CI) members and background (Bg) objects identification determined as explained in sect. 4.1.\\
Column 9: HI flux, in Jy km s$^{-1}$ corrected for pointing offset and beam attenuation consistently with Springob et al. (2005).\\
Column 10: HI line width at 50\%, defined as in Springob et al. (2005; $W_{F50}$), corrected for smoothing, redshift 
stretch and turbulent motions as prescribed by Springob et al. (2005).\\ 
Column 11: logarithm of the HI mass, in solar units.\\
Column 12: reference to the HI data.\\
Column 13: the HI deficiency parameter, determined as described in the text.\\
Column 14: a code for the HI line profile, from Springob et al. (2005) whenever available, or with the following criteria: 
1 or high signal to noise, two horn profiles, 2 for  
high signal to noise one horn profiles, 3 for fair profiles, 4 for low signal to noise, bad quality profiles and 5 for unseen profiles.\\

\begin{table*}
\caption{Galaxies with HI data}
\label{Tabdata}
{\scriptsize
\[
\begin{tabular}{cccccccccccccc}
\hline
\noalign{\smallskip}
Name         &  type                   &a	& b    &  mag	  &	 vel   &  Dist  &Memb&    SHI	& WHIc  & MHI	& ref$^a$&defHI&  Wqual\\
	     &			       &'	&'     & 	  & km s$^{-1}$&  Mpc	&    &Jykms$^{-1}$&km s$^{-1}$& M$\odot$&&     &       \\
\hline
NGC3712      &  SB?                    &1.70	& 0.60 &  14.90   &	 1580  &   - 	& Bg &    7.60  & 119	& 9.04  &   4	 &-0.07&    1\\
UGC6599      &  Ibm                    &1.20	& 0.70 &  16.50   &	 1569  &   - 	& Bg &    5.31  & 104	& 8.88  &   2	 &-0.12&    5\\
UGC6610      &  Scd:                   &2.10	& 0.40 &  15.00   &	 1851  &   - 	& Bg &    12.76 & 199	& 9.39  &   1	 &-0.20&    1\\
UGC6637      &  S0                     &0.90	& 0.46 &  15.00   &	 1836  &   - 	& Bg &    1.84  & 132	& 8.54  &   15   &-0.11&    5\\
UGC6684      &  IA?(s)mV:              &1.00	& 0.40 &  15.70   &	 1788  &   - 	& Bg &    5.78  & 124	& 9.02  &   2	 &-0.30&    1\\
IC2957       &  S0-a                   &0.60	& 0.40 &  14.80   &	 1776  &  8.6   & Bg &    2.24  & 88	& 8.60  &   1	 &-0.51&    3\\
UGC6782      &  ImV                    &2.00	& 2.00 &  15.00   &	 525   &   - 	& CI &    6.96  & 83	& 8.54  &   1	 & 0.21&    1\\
NGC3900      &  SA(r)0+                &3.16	& 1.70 &  12.50   &	 1798  &   - 	& Bg &    16.84 & 420	& 9.48  &   1	 &-0.10&    1\\
UGC6791      &  Scd:                   &2.05	& 1.41 &  14.99   &	 1852  &  29.3  & Bg &    5.10  & 212	& 8.99  &   1	 & 0.18&    1\\
NGC3899      &  SAB(s)b?pec            &2.45	& 1.07 &  13.21   &	 1779  &  10.4  & Bg &    4.87  & 159	& 8.94  &   2	 & 0.24&    2\\
UGC6881      &  Im:                    &1.40	& 0.70 &  17.00   &	 607   &   - 	& CI &    2.42  & 66	& 8.08  &   2	 & 0.45&    2\\
UGC6900      &  Sd                     &2.00	& 1.29 &  14.80   &	 590   &   - 	& CI &    1.69  & 83	& 7.92  &   2	 & 0.82&    1\\
NGC4020      &  Sbd?sp                 &2.00	& 0.73 &  13.28   &	 760   &  14.2  & CI &    14.64 & 161	& 8.84  &   1	 &-0.11&    1\\
CGCG157080   &  Irr                    &0.58	& 0.42 &  15.00   &	 714   &   - 	& CI &    2.64  & 125	& 8.12  &   3	 &-0.12&    2\\
NGC4032      &  Im:                    &1.90	& 1.80 &  12.81   &	 1268  &   - 	& CI &    22.10 &  94	& 9.04  &   5	 &-0.33&    2\\
UGC7007      &  SB(s)mIV               &0.84	& 0.79 &  15.50   &	 774   &   - 	& CI &    2.90  &  36	& 8.16  &   2	 & 0.06&    2\\
NGC4062      &  SA(s)c                 &1.59	& 1.16 &  12.50   &	 769   &  25.5  & CI &    24.13 & 293	& 9.57  &   1	 &-0.54&    1\\
NGC4080      &  Im?                    &1.20	& 0.50 &  14.28   &	 567   &  10.9  & CI &    2.86  & 138	& 7.90  &   2	 & 0.36&    1\\
IC2992       &  S0?                    &0.60	& 0.60 &  14.80   &	 580   &   - 	& CI &    1.42  & -	& 7.85  &   15   &-0.25&    5\\
UGC7131      &  Sdm:                   &1.50	& 0.40 &  15.10   &	 253   &   - 	& CI &    4.56  & 100	& 8.35  &   2	 & 0.22&    1\\
NGC4136      &  SAB(r)c                &4.00	& 3.70 &  11.69   &	 609   &   - 	& CI &    39.51 & 84	& 9.29  &   6	 & 0.00&    1\\
NGC4173      &  Sdm:                   &5.04	& 0.69 &  13.59   &	 1127  &   - 	& CI &    42.12 & 75	& 9.32  &   7	 &-0.01&    1\\
UGC7236      &  Im:                    &1.10	& 0.90 &  16.28   &	 945   &   - 	& CI &    4.75  & 57	& 8.37  &   8	 & 0.01&    4\\
NGC4203      &  SAB0-                  &3.40	& 3.20 &  11.80   &	 1086  &   - 	& CI &    48.11 & 234	& 9.38  &   9	 &-0.44&    1\\
NGC4204      &  SB(s)dm                &3.60	& 2.90 &  12.90   &	 856   &   - 	& CI &    34.28 & 65	& 9.23  &   7	 &-0.13&    1\\
MRK1315      &  BCD                    &1.05	& 0.60 &  16.50   &	 847   &   - 	& CI &    20.89 & -	& 9.02  &   15   &-0.66&    5\\
UGC7300      &  Im                     &1.40	& 1.20 &  14.90   &	 1210  &   - 	& CI &    14.51 & 77	& 8.86  &   1	 &-0.33&    1\\
UGC7321      &  Sd                     &6.74	& 0.43 &  14.00   &	 408   &  21.7  & CI &    44.51 & 217	& 9.69  &   1	 & 0.04&    1\\
NGC4245      &  SB(r)0/a:              &2.90	& 2.20 &  12.31   &	 815   &  15.3  & CI &    0.76  & 214	& 7.62  &   1	 & 1.23&    1\\
NGC4274      &  (R)SB(r)ab             &6.80	& 2.50 &  11.34   &	 930   &  17.1  & CI &    9.91  & 432	& 8.83  &   10   & 0.72&    1\\
NGC4278      &  E1-2                   &4.10	& 3.80 &  11.20   &	 649   &12.6$^b$& CI &    10.52 & 395	& 8.59  &   16   & 0.38&    2\\
NGC4286      &  SA(r)0/a:              &1.60	& 1.00 &  14.10   &	 644   &   - 	& CI &    0.71  & -	& 7.55  &   15   & 0.81&    5\\
UGC7428      &  Im:                    &1.30	& 1.20 &  14.10   &	 1137  &   - 	& CI &    7.35  & 49	& 8.56  &   2	 &-0.08&    2\\
IC3215       &  Sdm:                   &1.26	& 0.35 &  15.00   &	 1019  &   - 	& CI &    2.64  & 102	& 8.12  &   2	 & 0.35&    2\\
NGC4310      &  (R')SAB(r)0+?          &2.20	& 1.20 &  13.22   &	 913   &  10.4  & CI &    1.15  & 166	& 7.47  &   11   & 0.88&    2\\
NGC4314	     &	SB(rs)a 	       &4.20	& 3.70 &  11.43   &	 963   &   - 	& CI &    0.40  & 111	& 7.30  &   1	 & 1.92&    1\\
IC3247	     &	Sd      	       &1.86	& 0.26 &  15.25   &	 569   &  23.0  & CI &    3.50  & 152	& 8.64  &   1	 & 0.34&    1\\
NGC4359	     &	SB(rs)c?sp             &3.50	& 0.80 &  13.40   &	 1253  &   - 	& CI &    14.72 & -	& 8.86  &   15   & 0.33&    0\\
IC3308	     &	Sdm:    	       &1.30	& 0.20 &  15.41   &	 316   &   - 	& CI &    13.30 & 131	& 8.82  &   3	 &-0.34&    2\\
NGC4395	     &	SA(s)m: 	       &13.20	& 11.00&  10.64   &	 319   &  13.9  & CI &    401.61& 117	& 10.26 &   1	 &-0.40&    1\\
NGC4393	     &	SABd    	       &3.20	& 3.00 &  12.70   &	 755   &   - 	& CI &    25.35 & 92	& 9.10  &   15   &-0.07&    5\\
NGC4414	     &	SA(rs)c?	       &3.60	& 2.00 &  10.96   &	 716   &19.1$^b$& CI &    70.20 & 347	& 9.78  &   1	 &-0.36&    1\\
UGC7584	     &	Sdm:    	       &0.90	& 0.50 &  16.00   &	 603   &   - 	& CI &    3.65  & 33	& 8.26  &   1	 & 0.00&    1\\
NGC4448	     &	SB(r)ab 	       &3.90	& 1.40 &  12.00   &	 661   &  18.4  & CI &    3.00  & 377	& 8.38  &   1	 & 0.93&    1\\
NGC4455	     &	SB(s)d?sp	       &2.80	& 0.80 &  12.93   &	 637   &   - 	& CI &    30.67 & 98	& 9.18  &   1	 &-0.23&    1\\
UGC7673	     &	ImIII-IV	       &1.40	& 1.30 &  15.28   &	 642   &   - 	& CI &    10.38 & 73	& 8.71  &   1	 &-0.18&    1\\
UGC7698	     &	Im      	       &6.50	& 4.50 &  13.00   &	 331   &   - 	& CI &    40.98 & 59	& 9.31  &   1	 & 0.15&    1\\
NGC4509	     &	Sab pec?      	       &0.90	& 0.60 &  14.10   &	 937   &  6.1   & CI &    6.49  & 57	& 7.81  &   9	 & 0.21&    2\\
NGC4525	     &	Scd:    	       &2.37	& 0.96 &  13.40   &	 1172  &  9.2   & CI &    9.94  & 141	& 8.29  &   1	 & 0.28&    1\\
NGC4562	     &	SB(s)dm:sp             &2.32	& 0.69 &  13.90   &	 1353  &   - 	& CI &    8.29  & 121	& 8.61  &   1	 & 0.22&    1\\
NGC4559	     &	SAB(rs)cd	       &10.70	& 4.40 &  10.46   &	 816   &  7.2   & CI &    340.70& 236	& 9.62  &   1	 &-0.28&    1\\
IC3571	     &	BCD     	       &0.60	& 0.50 &  17.79   &	 1260  &   - 	& CI &    0.91  & -	& 7.65  &   17   & 0.36&    5\\
NGC4565	     &	SA(s)b?sp	       &15.90	& 1.85 &  10.42   &	 1230  &13.4$^b$& CI &    341.11& 394	& 10.16 &   1	 &-0.12&    3\\
UGCA292	     &	ImIV-V  	       &1.00	& 0.70 &  16.00   &	 308   &   - 	& CI &    15.64 & 28	& 8.89  &   1	 &-0.57&    1\\
NGC4631	     &	SB(s)d  	       &15.50	& 2.70 &   9.75   &	 606   &  5.3   & CI &   1077.60& 276	& 9.86  &   12   &-0.48&    1\\
MCG+06-28-022&	BCD		       &0.68	& 0.58 &  15.40   &	 892   &   - 	& CI &    1.42  & -	& 7.85  &   15   & 0.24&    5\\
NGC4656	     &	SB(s)m pec    	       &6.46	& 0.66 &  10.60   &	 664   &   - 	& CI &    190.55& -	& 9.98  &   15   &-0.52&    5\\
UGCA294	     &	S0 pec	    	       &0.70	& 0.40 &  15.10   &	 947   &   - 	& CI &    6.89  & 84	& 8.53  &   3	 &-0.82&    2\\
NGC4670	     &	SB(s)0/a pec:BCD       &1.40	& 1.10 &  13.09   &	 1069  &  20.9  & CI &   15.35  & 121	& 8.88  &   13   &-0.35&    1\\
UGCA298	     &	BCD?    	       &0.50	& 0.30 &  15.50   &	 801   &  1.9   & CI &    0.73  & 38	& 7.56  &   14   & 0.34&    2\\
NGC4725	     &	SAB(r)ab pec	       &10.70	& 7.60 &  10.11   &	 1206  &13.0$^b$& CI &    141.19& 382	& 9.75  &   1	 &-0.10&    1\\
UGC7990	     &	Im	  	       &1.00	& 0.60 &  17.00   &	 512   &   - 	& CI &    2.38  & 70	& 8.07  &   2	 & 0.25&    1\\
KUG1249+263  &	Irr		       &0.60	& 0.40 &  16.72   &	 1225  &   - 	& CI &    2.33  & -	& 8.06  &   15   &-0.05&    5\\
NGC4747	     &	Sbcd? pec sp  	       &3.50	& 1.20 &  12.96   &	 1190  &  10.7  & CI &    39.00 & 155	& 9.02  &   8	 &-0.12&    3\\
IC3840	     &	Sm/Irr: 	       &1.00	& 0.24 &  16.90   &	 583   &   - 	& CI &    2.23  & -	& 8.04  &   15   & 0.28&    5\\
UGC8011	     &	Im      	       &1.40	& 1.10 &  18.00   &	 776   &   - 	& CI &    10.18 & 116	& 8.70  &   1	 &-0.17&    1\\
MRK1338	     &	E	    	       &0.40	& 0.20 &  15.50   &	 1069  &   - 	& CI &    0.68  & 19	& 7.53  &   14   &-0.24&    2\\
NGC4789A     &	IB(s)mIV-V             &3.00	& 2.20 &  13.94   &	 374   &  19.7  & CI &    31.77 & 70	& 9.46  &   1	 &-0.28&    1\\
UGC8030	     &	Im:     	       &0.83	& 0.44 &  18.00   &	 628   &   - 	& CI &    1.36  & 34	& 7.83  &   15   & 0.38&    2\\
NGC4826	     &	(R)SA(rs)ab            &10.00	& 5.40 &   9.36   &	 408   &  4.9   & CI &    51.25 & 308	& 8.47  &   1	 & 0.65&    1\\
UGC8181	     &	Sdm:    	       &1.50	& 0.40 &  16.00   &	 886   &   - 	& CI &    3.99  & 80	& 8.30  &   2	 & 0.27&    1\\
UGC8333	     &	Im:     	       &1.30	& 0.40 &  17.00   &	 935   &   - 	& CI &    11.88 & 125	& 8.77  &   1	 &-0.29&    1\\																 \
\end{tabular}
\]
}
Notes: a: references to HI data:
1=Springob et al. 2005;
2=Schneider et al. (1990);
3=Scodeggio \& Gavazzi (1993);
4=Theureau et al. (1998);
5=Helou et al. (1984);
6=Lewis (1987);
7=Fisher \& Tully (1981);
8=Huchtmeier \& Richter (1989);
9=Noordermeer et al. (2005);
10=Huchtmeier (1982);
11=Chamaraux et al. (1987);
12=Rots (1980);
13=Thuan (1981);
14= Huchtmeier et al. (2005);
15=Hyperleda;
16=Burstein et al. (1987);
17=Dahlem et al. (2005)\\
b: from Ferrarese et al. (2000)
\end{table*}

\begin{table}
\caption{Coma I and Virgo clouds properties}
\label{Tabcomp}
{\scriptsize
\[
\begin{tabular}{cccccccc}
\hline
\noalign{\smallskip}
Cloud		& N.$^a$	& extension 		& S\%$^b$ 	&  $\sigma _{vel} (km s^{-1})$ 	& HI-def	\\
\hline
Coma I		& 115(55)	& $\sim$ 250$^{o2}$	& 68		&  307			& 0.06$\pm$0.44	\\ 
Virgo~W		& 173(37)	& $\sim$ 12$^{o2}$	& 54		&  437$^c$		& 0.63$\pm$0.68	\\
Virgo~M		& 39(23)	& $\sim$ 4$^{o2}$	& 62		&  666$^c$		& 0.24$\pm$0.57	\\	
\noalign{\smallskip}
\hline
\end{tabular}
\]
}
Notes: \\
a: number of cataloged objects to mag $\leq$ 18; the number in parenthesis gives the late-type objects with HI data. \\
b: spiral fraction to $M_B$ $\leq$ -15. \\
c: from Gavazzi et al. (1999).\\
\end{table}

\section{The derived parameters}

\subsection{Cloud membership}

\begin{figure}
   \centering
   \includegraphics[width=10cm,angle=0]{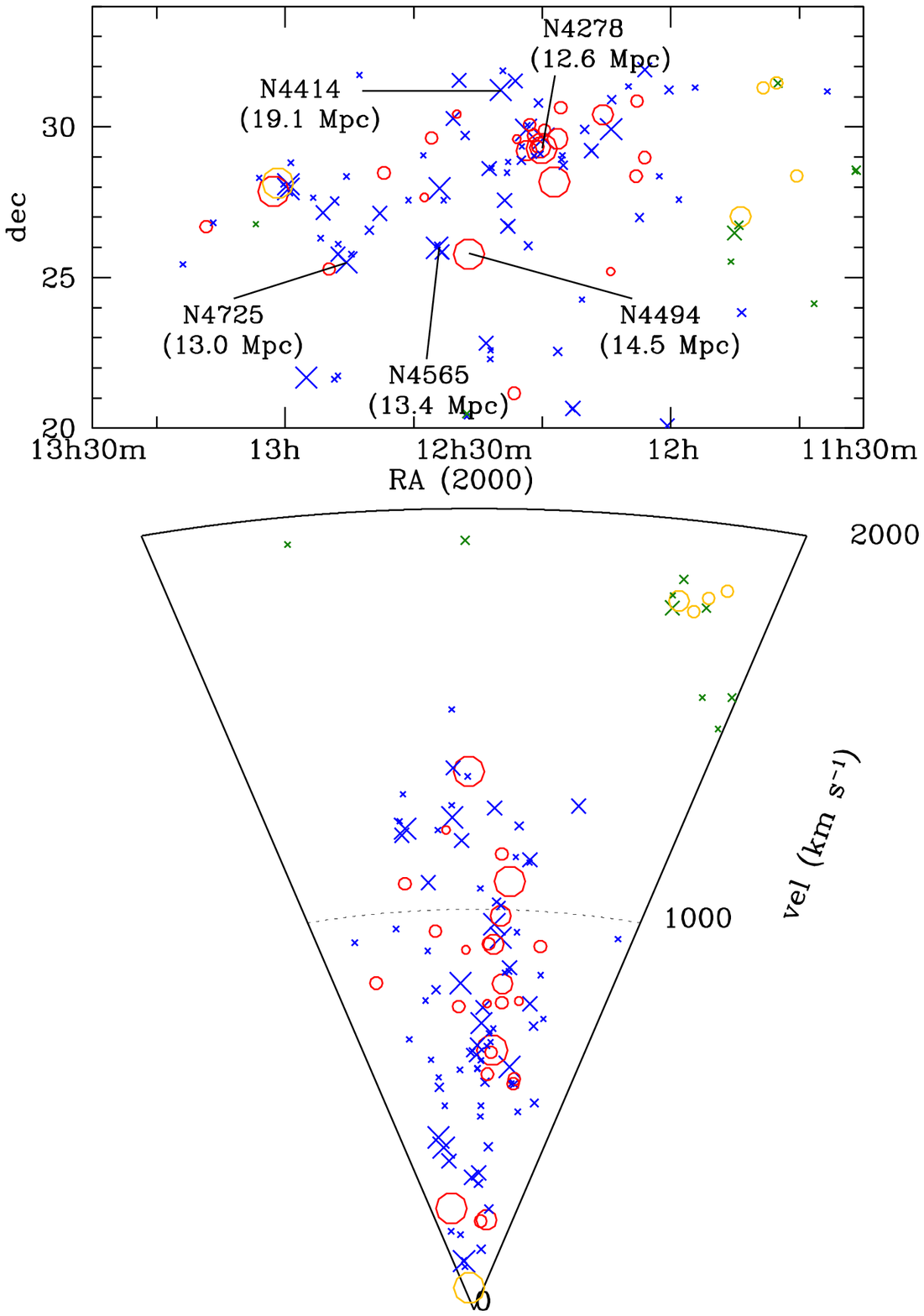}
   \caption{The sky distribution (upper panel) and the wedge diagram (lower panel) 
   of galaxies in the studied region. Red circles are for early-type ($\leq$ S0a), blue crosses
   for late-type galaxies in the Coma I cloud region as defined in the text, orange circles and green crosses
   for early- and late-type galaxies in the background ($vel_{hel}$ $>$ 1500 km s$^{-1}$) or in the foreground 
   ($vel_{hel}$ $<$ 100 km s$^{-1}$). The size of the symbols decreases from bright ($m_{pg}$ $<$ 12 mag) to weak 
   ($m_{pg}$ $\geq$ 16 mag) sources, in two bins magnitude intervals. }
   \label{type}
  \end{figure}

The distribution on the sky and on the velocity space of the selected galaxies
is shown in Fig. \ref{type}. The wedge diagram shown in the lower panel clearly
indicates that the Coma I cloud is confined within recessional velocities $\leq$ 1500 km s$^{-1}$.
To avoid any possible contamination from unclassified object maybe associated to our galaxy, we
arbitrarily remove any object with recessional velocity $<$ 100 km s$^{-1}$ (none of these objects have HI data).\\

The average recessional velocity of the galaxies identified as Coma I cloud members (132 objects
in the 100 $\leq$ $vel_{hel}$ $\leq$ 1500 km $^{-1}$)
is $vel_{hel}$ = 773 $\pm$ 307 km s$^{-1}$. The velocity dispersion is relatively small for a structure
extended over $\sim$ 250 square degrees in particular if compared to that observed in the
W and M clouds (see Table 2) located in the background of the Virgo cluster, at $\sim$ 32 Mpc (Gavazzi et al. 1999). \\
As defined, the Coma I cloud is an aggregation of objects with a spiral fraction of 68\%, thus
slightly higher than the W (54\%) and M (62\%) clouds in the Virgo cluster. 
An over dense region is visible around NGC 4278 ($\rho_{xyz}$ = 1.25 galaxies/Mpc$^3$, where $\rho_{xyz}$ is the local density 
of galaxies brighter than $M_B$ = -16, in galaxies/Mpc$^3$, within a three-dimensional grid 0.5 Mpc wide; Tully 1988b):
this density is slightly smaller than that observed in the periphery of the Virgo cluster 
($\rho_{xyz}$ $\sim$ 1.4 galaxies/Mpc$^3$; Tully 1988a).

\subsection{Distance determination}

An accurate determination of the gas properties of the target galaxies needs a precise distance determination. 
Five galaxies in the studied region have distance measurements from primary indicators
such as cepheids (NGC 4414, NGC 4725), Planetary Nebulae and Globular Cluster Luminosity Functions 
(NGC 4278, NGC 4494, NGC 4565) and surface brightness fluctuations (NGC 4278, NGC 4494, NGC 4565, NGC 4725)
(Ferrarese et al. 2000).\\
For inclined galaxies (inclination $\geq$ 30 deg) with available HI line widths and JHK total magnitudes,
the distance can be inferred using the Tully-Fisher relation determined adopting the Masters et al. (2008) calibration.
For these 27 galaxies we estimate their distance as the average of the JHK Tully-Fisher distance.  
We notice that for the three galaxies having both distance estimates, the Tully-Fisher distance 
is  systematically lower than that obtained from the primary indicators by $\sim$ 3.5 Mpc.\\
The average distance of the Coma I cloud members defined in the previous section\footnote{For galaxies with both primary distance
indicators and Tully-Fisher distances, the former are adopted.} is 13.91 $\pm$ 5.72 Mpc, while 14.52 $\pm$ 2.66 Mpc considering only
the five galaxies with primary distance indicators. In the following analysis we assume a distance of 14.52 Mpc for those galaxies
belonging to the Coma I cloud without any direct distance measurement. For the few background galaxies 
($vel_{hel}$ $\geq$ 1500 km s$^{-1}$)
the distance is determined assuming an Hubble constant of $H_0$ = 73 km s$^{-1}$ Mpc$^{-1}$
once their recessional velocity is corrected for a Virgocluster infall of 224 km s$^{-1}$.
 
 \begin{figure}
   \centering
   \includegraphics[width=10cm,angle=0]{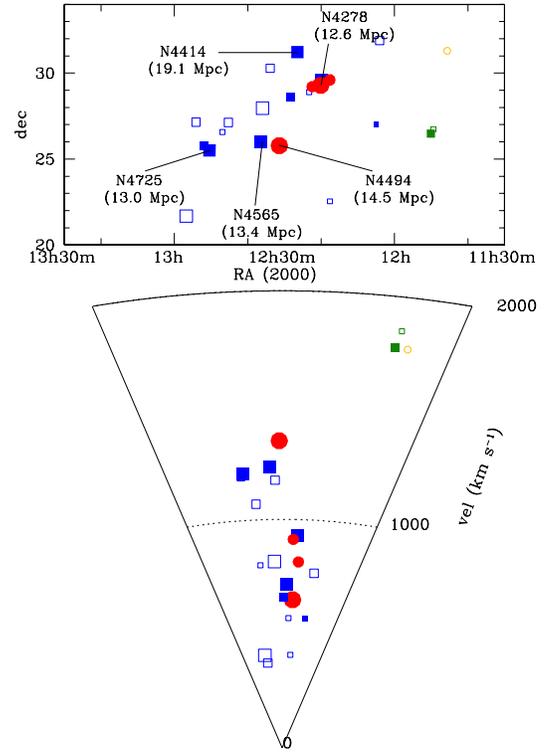}
   \caption{ Same as Fig. 1: filled symbols are for galaxies with a distance within 5 Mpc from the average 
   distance of the Coma I cloud as determined from primary indicators (14.52 $\pm$ 5 Mpc), empty symbols for galaxies outside this distance
   range. Red circles are for early-type ($\leq$ S0a), blue squares
   for late-type galaxies in the Coma I cloud, orange circles and green squares
   for early- and late-type galaxies in the background ($vel_{hel}$ $>$ 1500 km s$^{-1}$) or in the foreground 
   ($vel_{hel}$ $<$ 100 km s$^{-1}$).  
   The size of the symbols decreases from bright ($m_{pg}$ $<$ 12 mag) to weak 
   ($m_{pg}$ $\geq$ 16 mag) sources, in two bins magnitude intervals.
    }
   \label{dist}
  \end{figure}

\subsection{The gas mass and the HI-deficiency parameter}
 
The HI gas mass has been determined using the relation: 
\begin{equation}
{MHI = 2.36 \times 10^5 SHI(\rm Jy km s^{-1}) \it Dist^2(\rm Mpc)}
\end{equation}
where the distance is determined as described in the previous section.\\
The HI-deficiency parameter ($HI-def$) is defined
as the logarithmic difference between the average HI mass of a
reference sample of isolated galaxies of similar type and linear
dimension and the HI mass actually observed in individual objects:
$HI-def$ = Log $MHI_{ref}$ - Log$MHI_{obs}$. According to Haynes \&
Giovanelli (1984), Log$h^2 MHI_{ref}$ = $c$ + $d$Log$(h diam)^2$, where $c$
and $d$ are weak functions of the Hubble type, $diam$ is the linear
diameter of the galaxy (see Gavazzi et al. 2005) and $h$ = H$_0$/100.
We use in the present analysis the calibration of Solanes et al. (1996) for late-type
galaxies, extended to Scd-Im-BCD objects as prescribed in Gavazzi
et al. (in preparation) (see Table \ref{defcal}). 
This calibration is based on a sample of 98 galaxies of type $\geq$ Scd in the local supercluster (excluding the Virgo cluster),
observed by ALFALFA in the sky region 11$^h$ $<$ R.A.(2000) $<$ 16$^h$, 4$^o$ $<$ dec $<$ 16$^o$ and in
the velocity range 0$<$ $vel_{hel}$ $<$ 2000 km s$^{-1}$ and is, at present,
the best available calibration for this morphological class. It is preferred to the highly uncertain
calibration of Haynes \& Giovanelli (1984) which is based on a small sample of 38 Scd-Im-BCD
galaxies mostly of large diameter (Gavazzi et al. 2008; Solanes et al. 1996).

\begin{table}
\caption{The calibration of the HI deficiency parameter}
\label{defcal}
{\scriptsize
\[
\begin{tabular}{cccc}
\hline
\noalign{\smallskip}
Type	& c	& d	& Ref.\\
\hline
E-S0a	& 6.88	& 0.89	& HG84\\
Sa-Sab	& 7.75	& 0.59	& S96\\
Sb	& 7.82	& 0.62	& S96\\
Sbc	& 7.84	& 0.61	& S96\\
Sc	& 7.16 	& 0.87	& S96\\
Scd-Im-BCD & 7.45& 0.70	& G10\\
\noalign{\smallskip}
\hline
\end{tabular}
\]
}
References:\\
HG84: Haynes \& Giovanelli (1984); S96: Solanes et al. (1996); G10: Gavazzi et al. (in preparation).
\end{table}

 \begin{figure}
   \centering
   \includegraphics[width=10cm,angle=0]{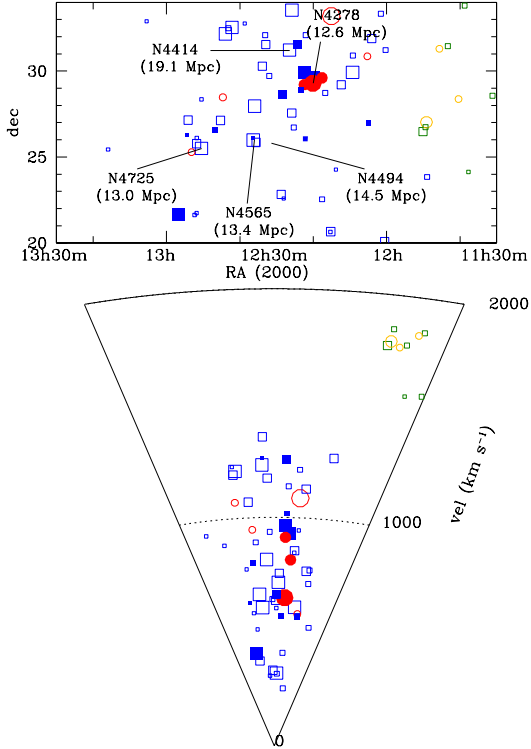}
   \caption{Same as Fig. 1: filled symbols are for HI-deficient galaxies ($HI-def$ $>$ 0.3), 
   empty symbols for galaxies with a normal HI content ($HI-def$ $\leq$ 0.3). Red circles are for early-type ($\leq$ S0a), blue squares
   for late-type galaxies in the Coma I cloud region as defined in the text, orange circles and green squares
   for early- and late-type galaxies in the background ($vel_{hel}$ $>$ 1500 km s$^{-1}$) or in the foreground 
   ($vel_{hel}$ $<$ 100 km s$^{-1}$). The size of the symbols decreases from bright ($m_{pg}$ $<$ 12 mag) to weak 
   ($m_{pg}$ $\geq$ 16 mag) sources, in two bins magnitude intervals. }
   \label{def}
  \end{figure}

\begin{figure}
   \centering
   \includegraphics[width=10cm,angle=0]{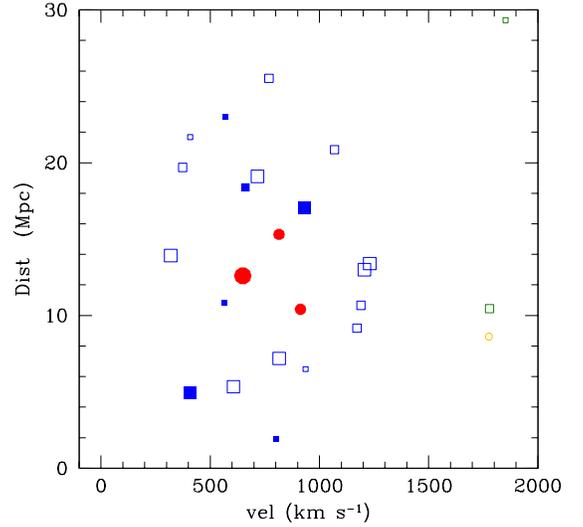}
   \caption{The distance-velocity diagram 
   of galaxies in the studied region. Filled symbols are for HI-deficient galaxies ($HI-def$ $>$ 0.3), 
   empty symbols for galaxies with a normal HI content ($HI-def$ $\leq$ 0.3). Red circles are for early-type ($\leq$ S0a), blue squares
   for late-type galaxies in the Coma I cloud region as defined in the text, orange circles and green squares
   for early- and late-type galaxies in the background ($vel_{hel}$ $>$ 1500 km s$^{-1}$) or in the foreground 
   ($vel_{hel}$ $<$ 100 km s$^{-1}$). The size of the symbols decreases from bright ($m_{pg}$ $<$ 12 mag) to weak 
   ($m_{pg}$ $\geq$ 16 mag) sources, in two bins magnitude intervals. 
    }
   \label{distvel}
  \end{figure}

\noindent
The average HI deficiency of galaxies in the Coma I cloud is $HI-def$ = 0.06 $\pm$ 0.44, thus slightly higher
than the average value for unperturbed field objects ($HI-def$ = 0.00 $\pm$ 0.30; Haynes \& Giovanelli 1984).
This result is robust against the adopted calibration of the HI-deficiency parameter since it does not change significantly 
using the $c$ = 7.00 and $d$ = 0.94 coefficients for Scd-Im-BCD galaxies of Haynes \& Giovanelli (1984): 
$HI-def$ = 0.03 $\pm$ 0.44, thus consistent with our estimate.   
Out of the 55 late-type Coma I cloud members with available HI data, only 13 (24 \%) can be considered as deficient in HI gas 
since having HI deficiencies greater than 0.3. These most deficient objects (filled squares in Fig. \ref{def}) do not seem to be located 
in  priviledged zones 
of the sky or of the velocity space, nor are objects at the average distance of the Coma I cloud but 
with high velocity with respect to the cloud (Fig. \ref{distvel}).   

\section{Discussion and conclusion}
 
 \begin{figure}
   \centering
   \includegraphics[width=10cm,angle=0]{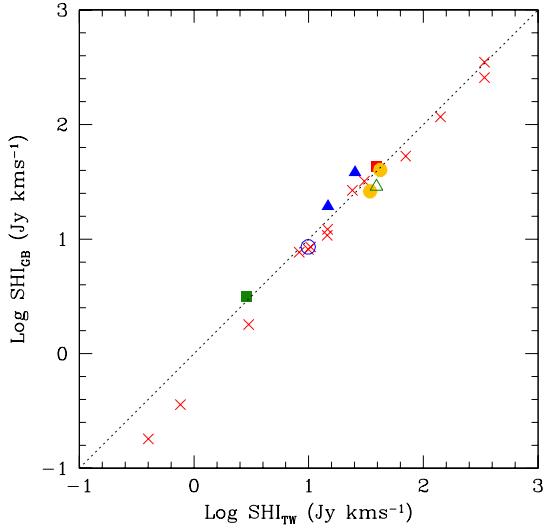}
   \caption{Comparison of the HI fluxes (in Jy km s$^{-1}$) used in this work to those determined from Table 1 
   of Garcia-Barreto et al. (1994) using a distance of 10 Mpc and the relation given in eq. (1). Red crosses indicates data from
   Springob et al. (2005), red filled squares from Lewis (1987), blue filled triangles from HyperLeda,
   green filled squares from Schneider et al. (1990), orange filled dots from Fisher \& Tully (1981), blue empty dots from Huchtmeier 
   (1982) and green empty triangles from Huchtmeier \& Richter (1989). The dotted line shows the one to one relation.
    }
   \label{comp}
  \end{figure}

By studying the HI properties of 32 galaxies in the Coma I cloud with data taken at Effelsberg, Garcia-Barreto et al. (1994) concluded that 
these objects are generally devoid of gas. In their sample of 23 late-type galaxies the average HI-deficiency is
0.40 $\pm$ 0.59, significantly higher than the value found in this work ($HI-def$=0.06 $\pm$ 0.44) on a more than
doubled sample (55 objects). 
The difference with Garcia-Barreto et al. (1994) might result from several reasons. 
The HI fluxes used in this work, mostly (64\%) taken from the compilation of Springob et al. (2005) 
are systematically higher (22\%) than those of Garcia-Barreto et al. (1994) (see Fig. \ref{comp}). 
The resulting HI-deficiency parameter is thus lower by a factor of 0.09 on average than the previous estimate.
This difference can be due to the fact that Springob et al. (2005) correct the data for beam attenuation and pointing offsets, 
while it is unclear whether Garcia-Barreto et al. (1994) used similar corrections.
An additional difference of 0.03 in $HI-def$ is due to the fact that, to transform $SHI$ fluxes into gas masses, Garcia-Barreto et al (1994)
used in eq. (1) a constant value of 2.22 $\times$ 10$^5$ instead of 2.36 $\times$ 10$^5$ as in this work.
The relationship between optical linear diameters and the HI mass being non linear, the HI-deficiency parameter is not a distance independent
value: for a given galaxy the HI-deficiency increases if its distance decreases. Garcia-Barreto et al. (1994) used a distance of
10 Mpc in the determination of the HI mass of their sample, while we used the Tully-Fisher distance whenever available, or
14.52 Mpc elsewhere. This difference in distance leads to an overestimate of the HI-deficiency parameter of $\sim$ 0.04
for a typical Sc galaxy in the Garcia-Barreto et al. calculations with respect to ours.
Conversely, the use of the calibration of Solanes et al. (1996) for Sa-Sc galaxies, which 
is based on $H_0$ =100 km s$^{-1}$ Mpc$^{-1}$, induces a decrease of the HI-deficiency parameter 
by a factor (1-$d$)Log$h^2$ (from 0.11 for Sa to 0.04 for Sc). 
Since the present sample is dominated by galaxies of type $\geq$ Scd (78\%), whose distance has been 
determined using $H_0$ =73 km s$^{-1}$ Mpc$^{-1}$, the average $HI-def$ is only 
marginally affected by the choice of $H_0$ =100 km s$^{-1}$ Mpc$^{-1}$ for Sa-Sc galaxies of Solanes et al. (1996).
The rest of the difference (0.18 in $HI-def$) might be due to statistical reasons, our sample (55 objects) being more than doubled with respect 
to that of Garcia-Barreto et al. (1994) (23 objects), or to the adopted calibration. Garcia-Barreto et al. (1994) determined the 
HI-deficiency parameter using the B band luminosity relation of Giovanelli et al. (1981), while our estimate is
based on a diameter relation. The calibration of the HI-deficiency on optical diameters
is less dispersed than that based on optical luminosities (Haynes \& Giovanelli 1984). \\
We can thus conclude that late-type galaxies in the Coma I cloud are not as deficient in HI gas as previously claimed. 
The Coma I cloud is thus composed of galaxies with a similar spiral fraction but richer in gas content than the Virgo M and W clouds. 
Being at a distance along the line of sight similar to that of Virgo (14.52 Mpc for Coma I and 16.5 Mpc for Virgo),  
and at a distance of $\sim$ 5 Mpc on the plane of the sky to the core of the cluster, it could be considered as a cloud of Virgo  
(for comparison the M and W clouds are located at $\sim$ 16 Mpc from the core of Virgo, Gavazzi et al. 1999). 
Is pre-processing acting on the late-type galaxy population inhabiting the Coma I cloud?
From a statistical point of view the present analysis excludes it. 
There existe however a fraction of objects with a significant HI-deficiency ($HI-def$ $>$ 0.3). What is their origin?
Because of the relatively poor statistics and the low density contrast within the cloud, it is impossible to disentangle gravitational interactions 
from interactions with the intergalactic medium within the Coma I cloud itself or during the crossing of the whole cloud through the core of the Virgo cluster.
The spread of the HI-deficient galaxies ($HI-def$ $>$ 0.3) within the cloud and in the velocity-distance space (Fig. \ref{distvel})
do not seem to favor the former scenario, since gravitational interactions or ram-pressure stripping within the cloud would 
be more efficient in the highest density regions or for galaxies with the highest velocities with respect to the mean value of the Coma I cloud.
Indeed using the prescription of Boselli \& Gavazzi (2006) we can estimate that the frequency of galaxy encounters within
the Coma I cloud is very low, the relaxation time being $\sim$ 40 Gyr.
Despite the process in place, however, if the Coma I cloud is representative of infalling groups in nearby clusters,
we conclude that in the nearby universe the gas properties of late-type galaxies belonging 
to large substractures of rich clusters do not appear significantly perturbed by their environment.
The undergoing ALFALFA survey (Giovanelli et al. 2005) will soon provide us with an unprecedent sky coverage in HI
of 7000 sq. degrees of the sky, thus covering a large range in galaxy density from the core of rich clusters to the local voids.
In particular, given its sensitivity (2.4 mJy at 5 km s$^{-1}$, Giovanelli et al. 2005) combined with a multi-beam detector, ALFALFA will be 
perfectly suited for observing at the same time extended sources and point-like objects as those populating the Coma I cloud.
This survey will thus be a unique opportunity for studying, using an homogeneous dataset and with an unprecedent statistical significance, 
the gas properties of galaxies in different density regimes of the local universe,
including loose groups and substructures probably infalling into rich clusters.
\\

\begin{acknowledgements}
      
We want to thank C. Marinoni, L. Cortese, C. Pacifici and S. Boissier for interesting discussions,
and the anonymous referee for useful comments. 
This research has made use of the NASA/IPAC Extragalactic Database (NED) 
which is operated by the Jet Propulsion Laboratory, California Institute of 
Technology, under contract with the National Aeronautics and Space Administration. 
We acknowledge the usage of the HyperLeda database (http://leda.univ-lyon1.fr)
and the GOLDMine database (http://goldmine.mib.infn.it/).

\end{acknowledgements}


\begin{thebibliography}{}


    
\bibitem[]{}Boselli, A. \& Gavazzi, G., 2006, PASP, 118, 517

%


\bibitem[]{}Boselli, A., Gavazzi, G., Donas, J., Scodeggio, M., 2001, AJ, 121, 753









\bibitem[]{}Burstein, D., Krumm, N., Salpeter, E., 1987, AJ, 94, 883

\bibitem[]{}Chamaraux P, Balkowski C., Fontanelli P., 1987, A\&AS, 69, 263

\bibitem[]{}Cortese, L., Gavazzi, G., Boselli, A., Iglesias-Paramo, J., Carrasco, L., 2004, A\&A, 425, 429

\bibitem[Cortese et al.(2006)]{2006A&A...453..847C} Cortese, L., Gavazzi, G., Boselli, A., Franzetti, P., Kennicutt, R.~C., O'Neil, K., \& Sakai, S.\ 2006, \aap, 453, 847 

\bibitem[]{}Dahlem, M., Ehle, M., Ryder, S., Vlajic, M., Haynes, R., 2005, A\&A, 432, 475


\bibitem[]{}Donnelly, R., Forman, W., Jones, C., et al., 2001, ApJ, 562, 254

\bibitem[]{}Dressler, A., 1980, 236, 351

\bibitem[]{}Dressler, A., 2004, in "Clusters of Galaxies: Probes of Cosmological Structure and Galaxy Evolution", from the Carnegie Observatories Centennial Symposia. Published by Cambridge University Press, as part of the Carnegie Observatories Astrophysics Series. Edited by J.S. Mulchaey, A. Dressler, and A. Oemler, 2004, p. 206.


\bibitem[]{}Ferrarese, L., Mould, J., Kennicutt, R., et al., 2000, ApJ, 529, 745

\bibitem[]{}Ferrari, C., Maurogordato, S., Cappi, A., Benoist, C., 2003, A\&A, 399, 813

\bibitem[]{}Fisher, J.R., \& Tully, B., 1981, ApJS, 47, 139

\bibitem[]{}Garcia-Barreto, J.A., Downes, D., Huchtmeier, W., 1994, A\&A, 288, 705

\bibitem[]{}Gavazzi, G., Boselli, A., 1996, Astro. Lett. and Communications, 35, 1





\bibitem[]{}Gavazzi, G., Boselli, A., Scodeggio, M., Pierini, D., Belsole, E., 1999, MNRAS, 304, 595




\bibitem[]{}Gavazzi, G., Boselli, A., Pedotti, P., Gallazzi, A., Carrasco, L., 2002, A\&A, 396, 449


\bibitem[]{}Gavazzi, G., Cortese, L., Boselli, A., et al. 2003a, ApJ, 597, 210

\bibitem[]{}Gavazzi, G., Boselli, A., Donati, A., Franzetti, P. \& Scodeggio, M., 2003b, A\&A, 400, 451



\bibitem[]{}Gavazzi, G., Boselli, A., van Driel, W., O'Neil, K., 2005, A\&A, 429, 439

\bibitem[]{}Gavazzi, G., Boselli, A., Cortese, L., Arosio, I, Gallazzi, A., Pedotti, P., 2006a, A\&A, 446, 839

\bibitem[]{}Gavazzi, G., O'Neil, K., Boselli, A., van Driel, W., 2006b, A\&A, 449, 929

\bibitem[]{}Gavazzi, G., Giovanelli, R., Haynes, M., et al., 2008, A\&A, 482, 43

\bibitem[]{}Giovanelli, R., Chincarini, G., Haynes, M., 1981, ApJ, 247, 383

\bibitem[]{}Giovanelli, R., Haynes, M., Kent, B., et al., 2005, AJ, 130, 2598

\bibitem[]{}Gnedin, O., 2003, ApJ, 582, 141

\bibitem[]{}Gomez, P., Nichol, R., Miller, C., et al., 2003, ApJ, 584, 210

\bibitem[]{}Gregory, S., Thompson, Li., 1977, ApJ, 213, 345

\bibitem[]{}Jarrett, T., Chester, T., Cutri, R., Schneider, S., Huchra, J., 2003, AJ, 125, 525

\bibitem[]{}Haynes, M., Giovanelli, R., 1984, AJ, 89, 758

\bibitem[]{}Helou, G., Hoffman, G. , Salpeter, E., 1984, ApJS, 55, 433

\bibitem[]{}Huchtmeier, W., 1982, A\&A, 110, 121

\bibitem[]{}Huchtmeier, W., \& Richter, O., 1989, in "A General Catalog of HI Observations of Galaxies", ed. Springer-Verlag 

\bibitem[]{}Huchtmeier, W., Krishna, G., \& Petrosian, A., 2005, A\&A, 434, 887

\bibitem[]{}Lewis, B.M., 1987, ApJS, 63, 515

\bibitem[]{}Lewis, I., Balogh, M., De Propris, R., et al., 2002, MNRAS, 334, 673

\bibitem[]{}Masters, K., Springob, C., Huchra, J., 2008, AJ, 135, 1738

\bibitem[]{}Noordermeer E., van der Hulst J., Sancisi R., Swaters R., van Albada T., 2005, A\&A, 442, 137

\bibitem[]{}Rots, A., 1980, A\&AS, 41, 189

\bibitem[]{}Sakai, S., Kennicutt, R., van der Hulst, J., Moss, K., 2002, ApJ, 578, 842

\bibitem[]{}Schneider, S., Thuan, T., Magri, C., Wadjak, J., 1990, ApJS, 72, 245

\bibitem[]{}Scodeggio, M., \& Gavazzi, G., 1993, ApJ, 409, 110

\bibitem[]{}Springob, C., Haynes, M., Giovanelli, R., Kent, B., 2005, ApJS, 160, 149

\bibitem[]{}Solnes, J., Giovanelli, R., Haynes, M., 1996, ApJ, 461, 609

\bibitem[]{}Theureau, G., Bottinelli, L., Coudreau, N., et al., 1998, A\&AS, 130, 333

\bibitem[]{}Thuan, T., 1981, ApJ, 247, 823

\bibitem[]{}Tully, B., 1988a, AJ, 96, 73

\bibitem[]{}Tully, B., 1988b, "Nearby Galaxy Catalog", Cambridge University Press

\bibitem[]{}Whitmore, B., Gilmore, D., 1991, ApJ, 367, 64

\end{thebibliography}
\end{document}